\documentstyle[twocolumn,prl,aps,epsfig]{revtex}
\title{Distributions of forces and 'hydrodynamic clustering' in a shear thickened colloid.}
\author{John R. Melrose}
\address{Polymers \& Colloids group, 
Cavendish Laboratory, Madingley Road, Cambridge, UK. CB3 OHE \\
\vspace{0.1in}
\parbox{5.5in}{\rm
Sheared concentrated colloids with short range polymer coats
are examined via simulations. Distributions of force are found to be sums of 
exponentials. The 'hydrodynamic 
clustering' underlying the shear thickening effect is shown, in this system, to be a network of
percolating coat contacts, with coats both in compression and extension. The geometry and kinetics of this network are explored along with its relation to the bulk stress tensor. Particles experience strong fluctuations in force over epochs $\it{circa}$ 10\%  strain.  Density fluctuations up to glassy
volume fractions give rise to scattering peaks at low Q in the thickening regime. There is some commonality in the physics of this colloid particulate system and other granular media. \\
\\ Pacs Numbers: 83.50.-v, 47.20.Ft, 83.50.Qm, 83.70.Hq}
}
\input epsf
\begin{document}
\def\vf{\phi_{v}}
\maketitle

What is the physics of concentrated particles in flow ?  Detailed observation is hard, however, simulations of powders/granular media \cite{thornton}, \cite{radji} have provided much insight.  They
have shown exponential distributions of force and force-chains. The origin of the distribution of force has motivated much theory [3-6], although in the main for static systems.


This letter examines force transmission in steady states of sheared colloid systems.  
It concerns shear thickening, a dramatic effect in which, at high
shear rates, concentrated colloids develop high, and sometimes diverging, viscosities \cite{barnes}.
Understanding thickening is of considerable interest [8-12].  
The concept of 'hydrodynamic clustering' has been introduced to explain thickening, but its precise meaning has not been explored.  The main aim of this letter is to clarify 'hydrodynamic clustering' , but on route to this commonalities will be found with dry powders.


This work considers colloids of rigid particles suspended in a fluid. 
Particle motion is overdamped. The equation of motion is

\begin{equation} \bf{R}\cdot \bf{V} + \bf{F}_{C} + \bf{F}_{B} = 0  , \end{equation} 
where $\bf{V}, \bf{F}_{C}, \bf{F}_{B}$ are $6N$ velocity/angular-velocity and colloid and Brownian 
forces/torque vectors and $\bf{R}$ is a 
$6N \times 6N $ drag matrix of dissipative hydrodynamic interactions.

The results below are, in the  main, for a model which includes just the squeeze mode of hydrodynamic lubrication forces 
between near neighbours.  
For spheres $i$ and $j$ with velocities $\bf{v}_{i}$ and $\bf{v}_j$ this determines a force $\bf{f}_{ij}$ on particle $i$

\begin{equation}\bf{f}_{ij}= - \alpha(\it{h})(\bf{v}_i-\bf{v}_j)\cdot\bf{n}_{ij}\bf{n}_{ij} , \end{equation}
where $\bf{n}_{ij}$ is the unit vector from centre $i$ to $j$, $\alpha(h)$ the drag coefficient 
and $h$ is the gap between the sphere
surfaces.  $\bf{R}$ is made up of such pair terms.  In the absence of dissipations for shear modes, the particles are 'frictionless'.
Some simulations have also been performed that include the shear mode.

The algorithm for simulating Eq. (1) with pair terms is given in
\cite{ball&melrosetec}. The lubrication force opposes the relative motion of the particles.
At the high concentrations of interest it is argued \cite{ball&melrosetec}
that these terms will dominate the long ranged mobility matrix in shear flow.  
In the case of a hard sphere model the drag coefficient $\alpha(h)$ is given by the well known Reynolds formula
which diverges as $\it{1/h}$. The particles simulated here have crude models for short range polymer coats with both a conservative repulsive force
and a lubrication interaction modified from that of hard spheres.  The coats define a particle contact which is crucial to the physics discussed below.

Experiment \cite{bender&wagner},and simulation  \cite{jmhvrb}, \cite{phung&brady} establish that thickening is dominated by contributions from the lubrication forces. 
However, Reynolds lubrication is insufficient to drive strong 
thickening. 
Crude models of particle pairs shearing past each other \cite{marrucci}, lead 
to the conclusion that any divergence of bulk properties can only be as the logarithm 
of the inverse gap of closest approach $h_{min}$ . It was suggested \cite{marrucci} that N-body effects 
may be necessary to explain the stronger divergences of experiment.
  
However, simulations of spheres with Reynolds lubrication \cite{jmhvrb} \cite{phung&brady} 
do show only logarithmic divergence, much weaker than experiment \cite{frith}.  
The correct interpretation of the 
arguments in ref. \cite{marrucci}
is not that N-body effects are necessary, but that lubrication interactions stronger than Reynolds are necessary.
Lubrication interaction can be strongly enhanced by polymer coats on the particles \cite{fred&pincus}. 
On compression of the coats the lubrication divergence switches to a higher power of $(1/h)$ than Reynolds lubrication.  The strength of its lubrication coefficient, 
depends on its porosity (the smaller the pores,
the stronger the lubrication coefficient; see \cite{jmhvrb} for a plot of this interaction). The coat also has an elastic modulus. It was shown previously \cite{jmhvrb} that this model fits experimental data \cite{frith}.

Simple pair arguments \cite{tohwa} explain the onset of thickening. Thickening occurs once the bulk stress is sufficient to 
compress the coats against their conservative (spring) forces to a point at which the relaxation time of the coats, $\tau_c$ is longer than the inverse shear rate $1/\dot{\gamma}$. This is confirmed
by the simulations.

Although we can understand the onset of thickening at the 
pair level, there is also a dependence on volume fraction, with strong thickening occurring only for $\vf > 0.50$. Simulations do reveal many body effects.  Simulations \cite{bossis&brady} motivated a model of 
thickening by {\it{hydrodynamic clustering}}: agglomerates of particles
bound together by lubrication forces.  Assuming such clusters are rigid, explains enhanced stress.   However lubrication forces are only active under relative motion, so rigidity and agglomeration under lubrication are not obviously compatible. Indeed the 'clusters' are reported to be transient \cite{bossis&brady} and observed to rapidly disappear after cessation of flow \cite{buteraetal} - they exist in response to the stress. 

In the simulations below, the thickening regime when $\dot{\gamma}\tau_c >1 $ is identified with a network of coat contacts of mean coordination in the range 5-6. It is this network that constitutes 'hydrodynamic clustering'. The geometry and kinetics of this network will be explored in detail.

The simulated model has a coat thickness set at $0.01d$ and other parameters choosen to fit experiments on perspex particles \cite{frith}. The shear rate is defined as the dimensionless Peclet number. 
Figure 1 shows simulation data at volume fraction $\vf = 0.54$.  Below  $Pe=200$, the viscosity is close to flat and the
particles flow in an ordered array of strings aligned along the flow direction. The system jumps from ordered flow to disordered flow from $Pe=200$ to $Pe=500$ where it continues to be disordered and thicken at higher rates. This order-disorder transition during the initial stages of thickening is common for
mono-disperse spheres, however it is not sufficient nor necessary for thickening which is a feature of the disordered flow. 
We report elsewhere \cite{catherall&melrose} systems where the onset of thickening is from regimes of 
disorder and charged systems at $\vf < 0.50$ which show order disorder changes with minimal thickening.  In any case, this letter is a study of the steady state thickening effect within the disordered regime.  (The steady state should be contrasted with jamming in hard spheres \cite{jmrb1}.  A jamming at a critical shear stress can also occur for models with coat interactions if the springs have a maximal force)

Thickening is accompanied by large normal stress differences.  
In agreement with experiment, $N_1$ is negative \cite{laun} 
and the coat lubrication forces are the dominant contribution to the stress tensor. At $Pe=200$, $\dot{\gamma}\tau_c >1 $ and remains so up to {\it{circa}} $Pe=10^4$ where the spring force of the highly compressed coats grows rapidly and the divergence of the lubrication coefficient weakens, bringing $\dot{\gamma}\tau_c $ close to unity and lowering the thickening.

Figure 2 shows, for several shear rates, the probability distribution for the magnitude of the 
lubrication force on individual bonds.  Bonds are defined as Voronoi near neighbours.
Data is shown at four different shear rates. It was confirmed that the effects reported are independent of systems size above N=500 and independent of time step over a decade of variation. On the linear-log plot the distributions at high force are clearly seen to be exponential.

\begin{figure}
  \begin{center}
    \leavevmode
    \centerline{\psfig{figure=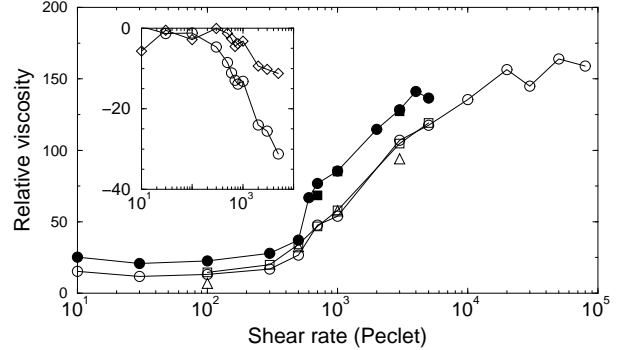,angle=270,width=8cm}}    
    \caption{Viscosity vs dimensionless shear rate (Peclet). Simulation at $\vf=0.54$ of colloid spheres with model polymer coats of thickness 0.01 diameters and porisity  $d_{p}= 0.0005 diam.$. Open symbols squeeze lubrication only, closed symbols with shear terms. Circles $N=50$ particles, squares $N=200$ and triangles $N=2000$. Inset normal stress differences $N_1/Pe$ (open circles) and $N_2/Pe$ (open diamonds).}
  \end{center}
\end{figure} 

In the  ordered phase at $Pe=100$ the distribution is a single exponential for the bonds in extension 
and dominated by a single exponential for the bonds in compression (there is negligible 2nd exponential in the tail). These are not observable on the scale of figure 2.

In the thickening regime at $Pe=500, 1000, 3000$  the distributions over the full range are fit by the sum of two exponential decays, one particular example is shown in the inset.  The system is characterised by two well separated characteristic forces. Bulk averages and the thickening are dominated by the growth of the high force distributions. Examination shows that the distribution with the decay at higher force (henceforth the {\it contact network}) is just that of the bonds with coats in contact; 45-50\% of the bonds are in the contact network; the mean number of contacts per particles is $5.2,5.7,5.9$ for  $Pe=500, 1000, 3000$ respectively. It is noted that 
the coordination number is approaching, from below, that of an isostatic network of frictionless particles \cite{ball&grinev}. 
In the ordered regime a percolating contact network also exists, but is of lower coordination  {\it{circa}} 3.

The finding of an exponential distribution of forces 
is common to recent simulations \cite{thornton} \cite{radji}and experiment \cite{meuth} on dry granular media. In shaken powders a bimodal distribution
was found \cite{radji}.  In sheared powders large fluctuations in normal stress were reported \cite{miller}. 
Note, we have also investigated force distributions when aggregation forces are present \cite{silbertetal} and for hard spheres with Reynolds lubrication and short repulsive springs. Both these show exponential distributions. The distribution is not the result of a particular force law.  Physical argument for this distribution have been given \cite{coppersmith}; it is geometric in origin.

\begin{figure}
  \begin{center}
    \leavevmode 
    \centerline{\psfig{figure=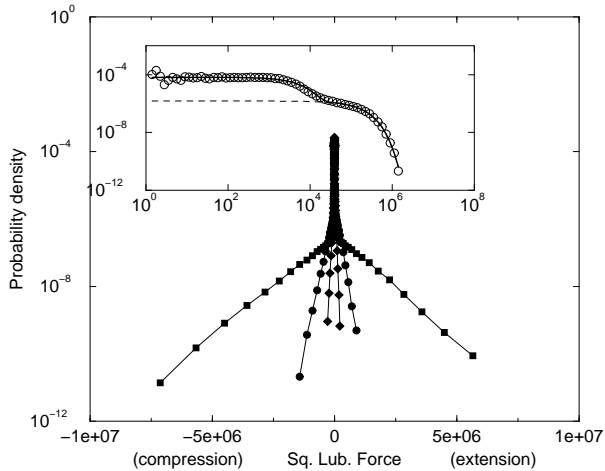,width=8cm}}        
    \caption{Probability density of forces (units kT/d) on bonds.  From the most inner pair to the most outer  pair $Pe=500, 1000 \,\,  \mbox{and}  \,\, 3000$ . Inset, a log-log plot of the distribution of force magnitudes for the compressive bonds at $Pe=1000$ (e.g.  the circles in the main plot 2nd in from the left).  Also shown in the inset is the fit of the whole distribution too a sum of two exponentials (solid line) and the fit of just the high force part (dashed line) to a single exponential. }
  \end{center}
\end{figure}

The geometry of the network was studied. Overall, the network is branched. By thresholding on force or interparticle gap, clusters of M particles were defined and their radius of gyration tensor examined.  At high force above the high characteristic forces, these had one eigenvalue scaling as M and others scaling roughly root M. Although linear at high force, they are not straight 'force-chains' \cite{cates}. By summing just over the contacts partial stress tensors, $\bf{\sigma}$, and a partial fabric tensor, F, were studied. 

\begin{equation} F=\sum_{ij} \bf{n}_{ij} \bf{n}_{ij} . \end{equation}

For the systems at Pe=500,1000 and 3000, table 1 gives data from the high force distributions.  
For the Pe=100 system it shows data for bonds 
with $ |f_{bond}| > 2 \times 10^3$. The flow geometry has y(x) the shear gradient(flow) direction with 
$(x=-y)$ the compression and $(x=y)$ the extension directions.
The table shows the  viscosity $\sigma_{xy}/Pe$ and first normal stress difference 
$N_1 = (\sigma_{xx}-\sigma_{yy})$, and the x and y components
of the normalised principal eigenvector of the fabric tensor.  The ordered flow at Pe=100 
has contacts in compression 
and extension just close to the gradient direction. For the thickened states contacts both  in compression and extension contribute to the shear stress. However, contacts in extension/compression contribute 
roughly 2/3 to 1/3 of the normal stress difference.  For contacts under compression, the fabric tensor has its principal axis just below the compression direction and is relatively independent of shear rate. For the bonds in extension its principal axis lies above the extension direction, thus  enhancing their contribution to $N_{1}$, it tends more to the extension direction the higher the shear rate. The lubrication forces are symmetric from extension to compression, but the spring forces are not.  It is the latter force that is responsible  for the asymmetry between the fabric of the compression and extension wings.   
   
\begin{center}
\begin{tabular}{|c|c|c|c|c|}
\hline
    Peclet Com./Ext.   &  N1/Pe     & visco    & x (flow) & y (grad.) \\
\hline    
    Pe=100  Com.       &  0.7       & 0.4      &  -0.43   & 0.9  \\
    Pe=100  Ext.       & -0.2       & 0.06     &   0.27   & 0.95 \\
    Pe=500  Com.       & -4         & 19       &  -0.75   & 0.65 \\      
    Pe=500  Ext.       & -6         & 9        &   0.58   & 0.81 \\  
    Pe=1000 Com.       & -5         & 34       &  -0.74   & 0.66 \\   
    Pe=1000 Ext.       & -12        & 19       &   0.62   & 0.78  \\
    Pe=3000 Com.       & -11        & 55       &  -0.74   & 0.68 \\  
    Pe=3000 Ext.       & -22        & 37       &   0.67   & 0.74  \\
\hline     
\end{tabular}
\end{center}
{Table 1 Contents are described in the text. }

\begin{figure}
  \begin{center}
    \leavevmode
    \centerline{\psfig{figure=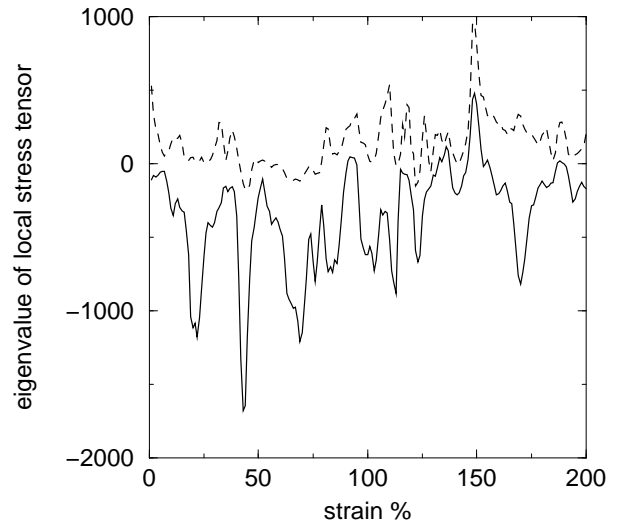,width=8cm}}    
    \caption{Short history of the most negative (compression) and most positive (extension) eigenvalues of
    the stress tensor summed just for the bonds surrounding one particular particle in the simulations at Pe=3000.
    The epoch shown is from strain 50 to 52.}
  \end{center}
\end{figure}

The kinetics of the networks was examined. It is straightforward to define 
an effective deformation tensor for a cluster.  'Rigid' clusters of particles defined on force or gap thresholds would have very low
extension and compression (relative to the applied shear) and just rotate in the flow.  No clear evidence of these was found, 
although some large clusters of contacts did have eigenvalues 
for the symmetric part of their deformation tensors as low as 1/4 of that for affine shear.  Figure 3 shows the large fluctuations in stress on a particle during flow; they build up and decay
over epochs of 10\% or so strain, at the applied shear rate this is much faster than the natural relaxation of the coat.  They are driven by the flow that is changing the geometrically determined stress concentration in the system. (In the frame co-rotating with the flow the systems experience a changing biaxial stress.)
Large changes in the distribution of forces under a 'small change' in the direction of applied stress has been termed 'fragility' \cite{cates}. 
It is unclear if the 10\% strain represents a 'small change' and whether the changes in the network are an example of 'fragility'.

A key question is whether the contact networks have an inherent length scale. The data
does suggest some correlation with density fluctuations. The structure factor in the thickening regime 
has peaks oriented along the flow direction inside the near neighbour ring. 
Similar peaks were observed in Neutron scattering 
from thickened colloids \cite{bender&wagner}.    
Selecting particles with high forces and 
computing a partial structure factor also resulted in low Q-peaks oriented along 
the flow direction.  Typically a particle must have candidate bonds for high forces close to the shear plane. If these are to carry high forces, and if the bonds through the particle are not co-linear, the other bonds must be able to supply large lateral forces to maintain overall balance.
A locally high density in which particles coats were compressed 
in all directions would provide this.    The distribution of
a local volume fraction parameter defined as a sphere volume divide by the volume 
of its Voronoi cell was examined. Over all particles this is peaked at 0.54. 
If the distribution is computed just over particles selected on a high threshold of force
magnitude, the peak is shifted to 
around local volume fractions 0.58-0.6, {\it circa} the hard sphere glass transition. Only to this degree can the network be considered 'clustering'. However, some particles at low local volume fractions can also carry high forces.

This work has detailed the nature of steady state thickening in colloids. Some features were common with the physics of slow powder flow - despite the different nature of the physical interactions.  Clearly this is due to the
generic importance of contact geometry in particle flows. Hopefully this will lead to exchanges of ideas and results. It is noted that the force network here does not conform to the simple notion of 'force chains' prevalent in much of the literature, it is branched and coordinated close to the iso-static limit, but very high force parts were like short segments of random walks. The existence and relevance of exponential distributions of force have been questioned, but figure 2 shows that, at least in flow, this is a significant feature.

For the colloid community, the results have given a much needed clarification and
detailing of the nature of 'hydrodynamic clustering'. There are no rigid clusters as some have assumed, but a network of contacts whose natural relaxation is slower than the shear time.   This network flows with large fluctuations in lubrication force driven by the changing geometry of contacts in the shear flow.  It had not been  realised that bonds both in compression and extension are involved.  The origin of normal stress differences is due to the fabric of the contact network.  Low Q density fluctuations observed in experiment are also associated with the contact network. The growth of the high force distribution and the percolating contact network could be taken as an 'order-parameter' for the thickening transition.

I acknowledge the support of the Unilever colloid physics program, EPSRC soft-solids Grant No GR/L21747
and funds from IFPRI . I have benefited from many useful discussions 
with R. Ball, R. Farr, L. Silbert, G. Soga, A. Catherall, D. Grinev, S. Edwards and M. Cates.

\end{document}